\documentclass[aps,prb,twocolumn]{revtex4}
\usepackage{graphicx}
\usepackage{amssymb,amsfonts,amsmath}
\begin{document}
\title{
Anomalous properties of the acoustic excitations in glasses 
on the mesoscopic length-scale}
\author{Giulio Monaco}
\email[]{gmonaco@esrf.fr}
\affiliation{European Synchrotron Radiation Facility, 6
rue Jules Horowitz, BP~220, 38043 Grenoble Cedex, France}
\author{Stefano Mossa}
\email[]{stefano.mossa@cea.fr}
\affiliation{UMR 5819 (UJF, CNRS, CEA) CEA, INAC, SPrAM,
17 Rue des Martyrs, 38054 Grenoble Cedex 9, France}
\affiliation{European Synchrotron Radiation Facility, 6
rue Jules Horowitz, BP~220, 38043 Grenoble Cedex, France}
\begin{abstract} 
The low-temperature thermal properties of dielectric crystals are 
governed by acoustic excitations with large wavelengths that are well 
described by plane waves. This is the Debye model, which rests 
on the assumption that the medium is an elastic continuum, holds 
true for acoustic wavelengths large on the microscopic scale fixed 
by the interatomic spacing, and gradually breaks down on approaching 
it. Glasses are characterized as well by universal low-temperature 
thermal properties, that are however anomalous with respect to 
those of the corresponding crystalline phases. Related universal 
anomalies also appear in the low-frequency vibrational density 
of states and, despite of a longstanding debate, still remain 
poorly understood.
Using molecular dynamics simulations of a model monatomic glass of 
extremely large size, we show that in glasses the structural disorder 
undermines the Debye model in a subtle way: the elastic continuum 
approximation for the acoustic excitations breaks down abruptly on 
the mesoscopic, medium-range-order length-scale of about ten interatomic 
spacings, where it still works well for the corresponding crystalline 
systems. On this scale, the sound velocity shows a marked reduction with 
respect to the macroscopic value. 
This turns out to be closely related to the universal excess 
over the Debye model prediction found in glasses at frequencies of 
$\sim$1~THz in the vibrational density of states or at temperatures 
of $\sim$10~K in the specific heat. 
\end{abstract}
\maketitle
\section{Introduction}
Glasses are structurally disordered systems.
As common experience shows, in the macroscopic limit they support sound 
waves as the corresponding crystalline materials do. In fact, averaging 
on a length-scale large enough, the details of the microscopic arrangement 
become essentially irrelevant. 
This holds true for sound waves with long wavelengths of at least several 
hundreds of nanometers, corresponding to wavenumbers, $q$, in the 
10$^{-2}$ nm$^{-1}$ range, as those probed with light scattering 
techniques~\cite{shapiro66}. 
On further increasing $q$, the effect of the structural disorder must at one 
point appear. 
The much larger $q$-scale of few nm$^{-1}$ is also well known, since it can be 
accessed experimentally with inelastic x-ray~\cite{sette98} and neutron~\cite{bove05} 
scattering techniques, and numerically with molecular dynamics 
simulations~\cite{rahman76,grest82,mazzacurati96,taraskin99,angelani00,ruocco00,horbach01,schober04}. 
These studies typically probe the dynamic structure factor, $S(q,\omega)$, that
is the space and time Fourier transform of the density-density correlation
function. They clearly indicate the existence in glasses of excitations that 
appear in those spectra as very broad peaks whose position as a function of $q$ 
is characterized by a sinusoidal-like dispersion curve.
Thus, these excitations strongly recall the acoustic modes in (poly-)crystalline 
systems up to roughly one half of the pseudo Brillouin zone~\cite{grest84}, i.e. 
down to distances corresponding to the interatomic spacing. For this reason, they 
are often dubbed as acoustic-like. However, as their broadening clearly indicates, 
they are far from being crystal-like modes and correspond in fact to a complex 
pattern of atomic motions~\cite{mazzacurati96,taraskin99,angelani00,ruocco00,horbach01,schober04}. 

Unfortunately, experimental and numerical studies leave a gap between 
the few nanometers scale and the hundreds of nanometers one that is extremely 
difficult to access. This keeps still open a number of fundamental questions 
in the physics of glasses and, more in general, on the nature of the vibrational 
excitations in disordered systems. 
{\em i)} How does the transition for the acoustic-like excitations look 
like between the small-$q$ Debye-like behaviour and the large-$q$ regime? 
In other words, how does it happen that reasonably well defined plane waves 
transform, on increasing $q$, into a complex pattern of atomic motions that 
mirror the structural disorder? 
{\em ii)} Similarly to crystalline systems, glasses are characterized by a 
universal behaviour in some fundamental low-temperature observables like 
specific heat and thermal conductivity~\cite{phillips81}. 
In particular, in the $\sim$10~K range the specific heat shows an excess with 
respect to the $T^3$, Debye-model prediction and the thermal conductivity shows 
a plateau. Despite of a longstanding 
debate~\cite{karpov83,akkermans85,dove97,schirmacher98,hehlen00,taraskin01,wittmer02,gurevich03,lubchenko03,grigera03,leonforte05,schirmacher06,schirmacher07,shintani08},
their origin is still poorly understood. It is however generally accepted 
that they are related to the ubiquitous existence at frequencies of $\sim$1~THz 
of an excess of modes in the vibrational density of states, $g(\omega)$, over 
the Debye-model prediction $g_D(\omega)= 3 \omega^2 / \omega_D^3$, 
with $\omega_D$ the Debye frequency.
This excess of modes is best visible in the reduced density of states, 
$g(\omega) / \omega^2$, where it appears as a broad feature known as boson
peak~\cite{buchenau84,malinovsky91}. 
Is then this universal behaviour in the low-temperature thermal properties 
or in the vibrational density of states related to the peculiar 
nature of the acoustic-like excitations in the low-frequency range? or, 
alternatively, do we have to imagine that glasses are characterized by additional 
low-frequency modes? 
{\em iii)} What is the information content that we can extract from 
the high-$q$ acoustic-like excitations measured in inelastic scattering 
experiments or calculated numerically? In which physical properties are their 
crystal-like features (e.g. existence of dispersion curves) reflected?  

The experiments that have attempted to answer these questions accessing the difficult 
10$^{-2}$--1~nm$^{-1}$ $q$-range lead to contrasting interpretations. An experiment 
based on a tunnel junction technique reported~\cite{rothenfusser83} linear dispersion 
for the transverse acoustic excitations in a silica glass up to frequencies 50\% 
smaller than the boson peak position in that glass ($\sim$1~THz). 
This result indicated that the acoustic excitations are unaffected in the frequency range 
relevant for the thermal properties in the $\sim$10~K range, and seemed to exclude 
any acoustic contribution to the low-temperature anomalies in the specific heat of 
glasses~\cite{rothenfusser83}. 
Early inelastic x-ray scattering results~\cite{sette98} seemed to 
confirm that scenario and showed crystal-like dispersing high-frequency longitudinal 
acoustic-like excitations with a broadening increasing quadratically with the
frequency. Again, a simple linear dispersion of the longitudinal acoustic-like modes 
was observed at frequencies corresponding to the boson peak position, thus suggesting 
a smooth transition between the macroscopic and microscopic regime. 
Recent and more accurate inelastic x-ray scattering studies have however revealed 
that the boson peak marks the energy where a qualitative change takes place: the 
longitudinal acoustic-like excitations show, below the boson peak position, a marked 
decrease of the phase velocity~\cite{monaco09} and a broadening characterized by a 
remarkable fourth-power-law frequency dependence~\cite{ruffle03,ruffle06,monaco09}. 
A similar behaviour for the broadening of the transverse acoustic modes had been 
found in a silica glass at low temperature and at frequencies below the boson 
peak position using a tunnel junction technique~\cite{dietsche79}. 
Conversely, a recent experiment using inelastic ultraviolet light scattering to 
measure the longitudinal acoustic modes in a silica glass at room temperature 
reported the onset of this peculiar $\omega^4$ regime at frequencies 
one order of magnitude smaller than the boson peak position~\cite{masciovecchio06}: 
the boson peak would then be not directly related to this regime. 
A complex and sometimes contradictory picture seems therefore to come out of the 
experiments performed so far.

Classical molecular dynamics simulations have provided a complementary tool 
to study the vibrational properties of glasses, starting from the pioneering 
investigations of Rahman and coworkers~\cite{rahman76}. The body of results available 
until now supports a scenario where the longitudinal acoustic-like excitations seem 
to be largely decoupled from the boson peak: they show a linear dispersion and a 
broadening that grows quadratically with frequency with no special feature in the 
frequency region where the boson peak 
appears~\cite{mazzacurati96,taraskin99,angelani00,ruocco00,horbach01,schober04}.
However, these studies could not provide a final answer on this issue due to the 
fact that the largest wavelengths that could be studied so far -- fixed by the 
simulation box size and then ultimately by computer power -- were still in the 
range of few tens of interatomic spacings. As a consequence, the corresponding 
lowest frequency acoustic-like excitations that were accessible lied too 
close to the boson peak position to allow for definite conclusions in this 
supposingly crucial frequency range.

We report here on a study of the vibrational dynamics of the classical Lennard-Jones 
(LJ) monatomic glass model~\cite{rahman76}. The reason of our choice is simple: despite 
of the fact that this system easily crystallizes below the melting temperature and thus 
requires extremely fast and experimentally out of reach quenching rates in order to prepare 
a glass starting from the melt, it is the simplest realization of a structural glass 
at our disposal. It thus provides the very basic ingredients of the vibrational 
dynamics of a structural glass that in other systems might be superimposed to additional 
more complex effects.  
Specifically, we present here results for an exceptionally large simulation box containing 
up to $N$=10$^7$ particles and clarify how the acoustic modes look like in the frequency 
region where 
the boson peak appears. In particular, we managed to probe acoustic excitations down 
to frequencies one order of magnitude lower than the boson peak position. Thanks to 
that, the study of their $q$-dependence allows us to establish that the boson peak 
originates from a deformation of the dispersion curves with respect to the crystal case. 
This boils down to a direct connection between the boson peak and the breakdown of 
the Debye continuum approximation for the acoustic excitations that takes place on a 
length-scale matching that of the medium range order of the glass~\cite{wittmer02,leonforte05,monaco09}.
\section{Results and discussion}
Details about computer simulation methods used are given in Appendix~\ref{appendix-methods}. 
We have calculated both the transverse, $S_T(q,\omega)$, and the longitudinal, 
$S_L(q,\omega)$, dynamic structure factors which give information on the transverse 
and longitudinal acoustic-like excitations, respectively. It is important to underline that 
while the latter can be obtained experimentally using scattering techniques, the former can 
only be studied in the frequency range relevant here using computer simulations. 
In what follows we will use LJ units. To make connection with experiments, 
we recall that, if we model argon using the LJ potential, then the temperature scale 
is in units of $\epsilon$=125.2~K, the length scale in units of $\sigma$=3.405~$\AA$,
and the time scale in units of $\tau$=2.11~ps.
In Fig.~1 some representative $S_T(q,\omega)$ spectra are reported, including 
those corresponding to the lowest $q$ value that we could reach in our simulation. 
\begin{figure}
\centering
\includegraphics[width=0.48\textwidth]{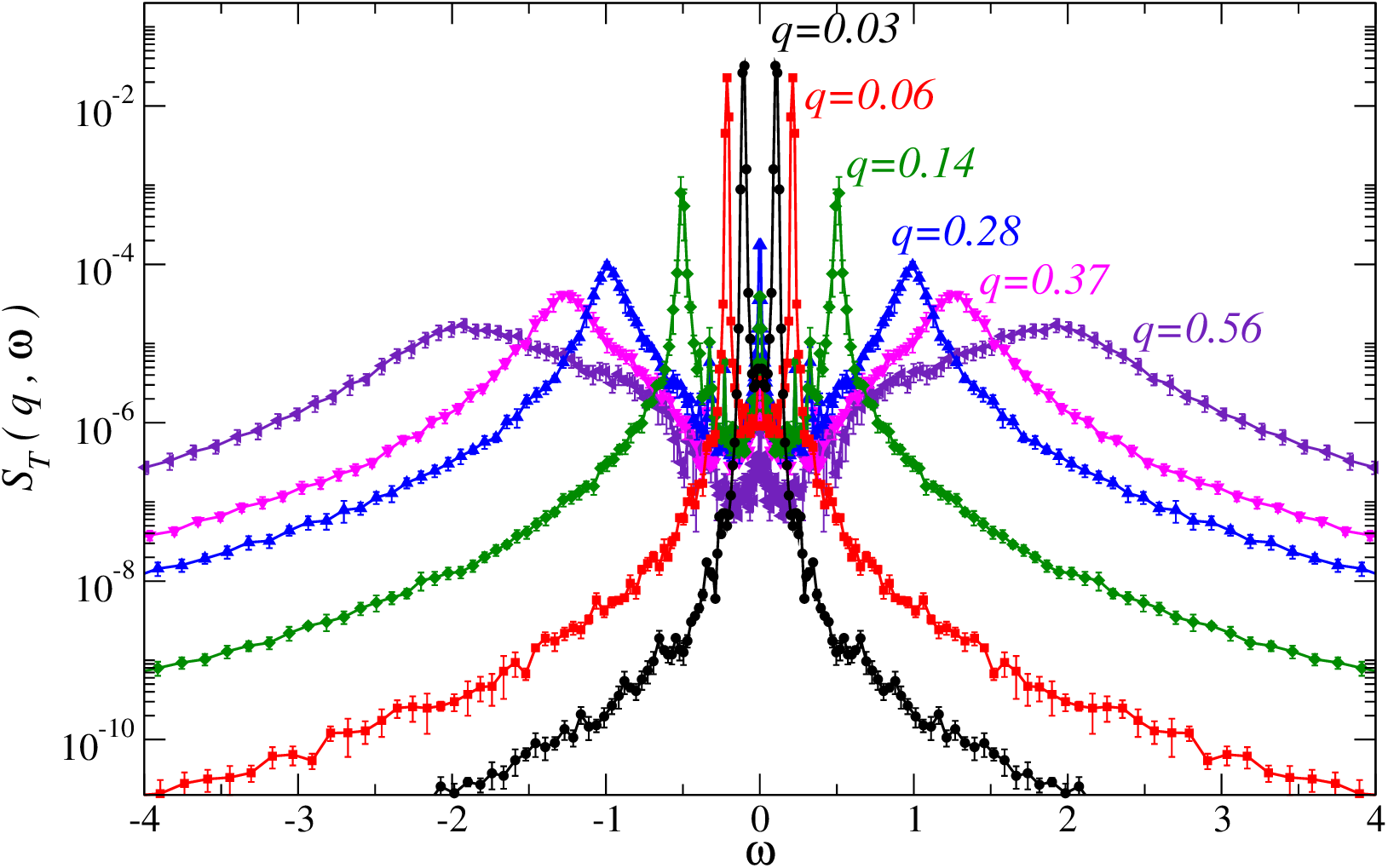}
\caption{
Transverse dynamical structure factors, $S_T(q,\omega)$, for a LJ glass 
at number density  $\widehat{\rho}$=1.015, temperature $T$=10$^{-3}$ and at 
the indicated $q$ values, including the smallest one accessible using a 
simulation box containing $\sim$10$^7$ atoms. The Brillouin peaks shift
towards higher frequencies and show a clear broadening on increasing $q$.
}
\label{Fig1}
\end{figure}  

For reference, we recall that at the studied number density $\hat{\rho}=N / V=$1.015 
(V being the simulation box volume) the melting temperature of the LJ system 
is $T_m \simeq$~1, and the glass-transition temperature $T_g \simeq$~0.4~\cite{robles03}. 
For what concerns the glass that we study, the Debye frequency and wavenumber are 
$\omega_D$=16.2 and $q_D$=3.92, respectively; the first sharp diffraction peak is at 
$q_m \simeq$~7, 
that corresponds to an average nearest neighbours distance of $\simeq 2 \pi /q_m =$0.9.
The spectra reported in Fig.~1 refer then to $q$-values down to $\sim$10$^2$ times 
smaller than the border of the pseudo-Brillouin zone located at $\simeq q_m /2$. 

The $S_T(q,\omega)$ spectra are characterized by two symmetric peaks (Brillouin 
peaks) in addition to a sharp elastic peak at $\omega=$0. The
Brillouin peaks can be characterized by the position of the maximum and the 
broadening. This information has been obtained by fitting a damped 
harmonic oscillator model to the spectral region, $I_{B}(q,\omega)$, around
the Brillouin peaks: 
\begin{equation}
I_{B}(q,\omega) \propto \frac{\Gamma_T(q) 
\Omega_T^2(q)}{(\omega^2-\Omega_T^2(q))^2 + \omega^2 \Gamma_T^2(q)}.
\label{Eq1}
\end{equation}
The parameters $\Omega_T(q)$ and $\Gamma_T(q)$ represent the characteristic 
frequency and broadening (FWHM) of the Brillouin peaks, respectively. 
The parameter $\Omega_T$ is utilized to obtain the transverse sound phase velocity, 
$c_T(q)=\Omega_T(q) / q$, and is reported in Fig.~2a as a function of 
frequency. 
\begin{figure}[b]
\centering
\includegraphics[width=0.48\textwidth]{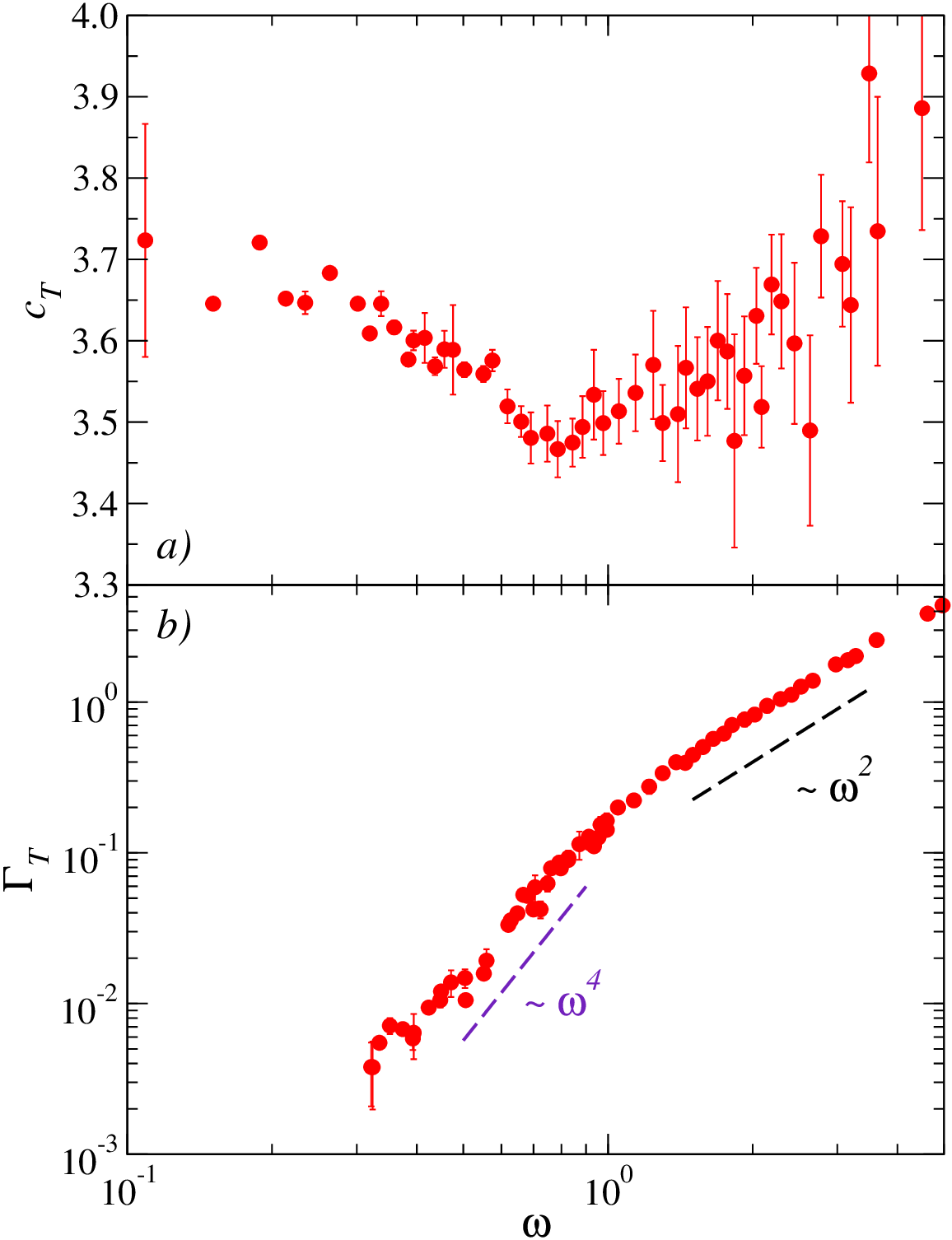}
\caption{
Phase velocity and broadening of the transverse acoustic-like excitations 
in a Lennard-Jones glass. 
Frequency dependence of phase velocity (red circles, panel a) and FWHM 
(red circles, panel b) of the transverse acoustic-like excitations of 
the studied LJ glass. The dashed lines in b) 
emphasize different regimes: $\sim \omega^2$ at high-frequencies and 
$\sim \omega^4$ at low-frequencies. The transition between the two regimes 
appears at the frequency where the phase velocity in a) shows a minimum. 
}
\label{Fig2}
\end{figure}

These data show an increase with frequency (positive dispersion) of the sound 
velocity for $\omega > 0.8$ as already reported for the longitudinal 
excitations in the same glass~\cite{ruocco00}. It is interesting to 
observe that below that frequency the macroscopic $\omega=$0 sound velocity limit 
is not directly recovered: instead a previously unnoticed region where the phase 
velocity decreases with increasing frequency (softening) appears. This is exactly 
the region where the boson peak is found in this LJ glass (see Fig.~4b below). 
The boson peak then appears not in a frequency range where the acoustic-like 
excitations disperse linearly (constant phase velocity), at odds with what was 
previously thought~\cite{rothenfusser83,sette98}, but where they experience a 
more complex dispersion behaviour. This is important because it clarifies that 
the Debye continuum approximation for the acoustic excitations breaks down at 
frequencies comparable to the boson peak position, i.e. at frequencies much lower 
than previously expected. 

Complementary information on this issue can be found in Fig.~2b, where the 
frequency-dependence of the broadening, $\Gamma_T$, of the transverse 
acoustic-like excitations is presented. These data clearly show two different regimes: 
a $\omega^2$ regime at high-frequencies, as already known from previous studies 
and associated to the structural disorder of the glass~\cite{taraskin99,ruocco00,horbach01}, 
and a previously unnoticed $\omega^4$ regime at low frequencies. The frequency that marks 
the change of regime is very close to, though slightly lower than, the boson peak position 
(see Fig.~4b). Moreover, the low frequency regime appears in the same range where the softening 
of the sound velocity shows up in Fig.~2a, thus indicating that these two features must 
be connected. This, however, is not surprising if one recalls that the Brillouin position and 
broadening can be related to the real and imaginary part of a complex self-energy, 
respectively~\cite{schirmacher06}.
\begin{figure}[t]
\centering
\includegraphics[width=0.48\textwidth]{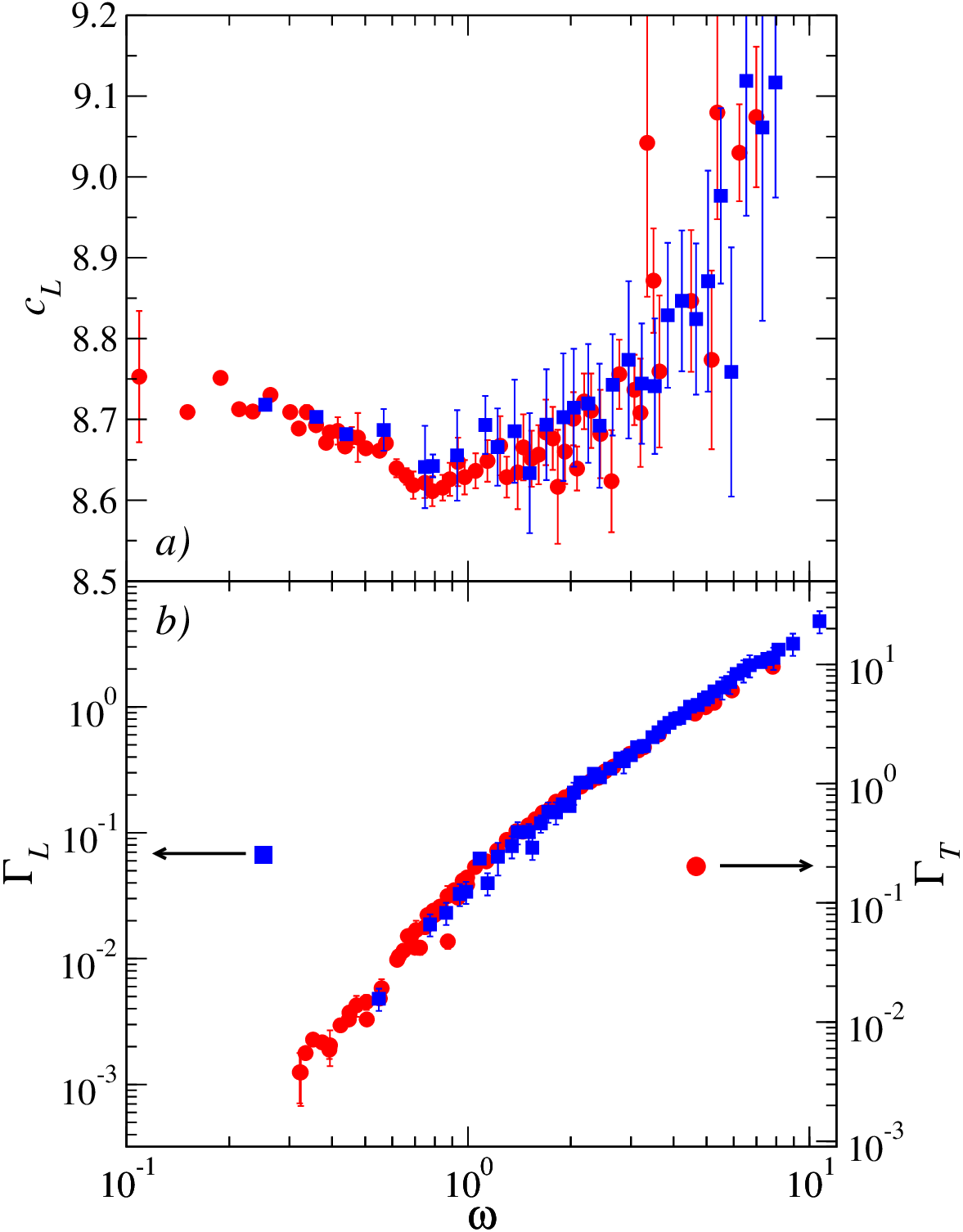}
\caption{
Comparison between transverse and longitudinal acoustic-like excitations 
in a Lennard-Jones glass.  Frequency dependence of phase velocity (blue squares, 
panel a) and broadening (blue squares, panel b, left axis) of the longitudinal 
acoustic-like excitations of the studied LJ glass. These data, similarly to those 
in Fig.~2, have been derived by fitting a damped harmonic oscillator model to 
the longitudinal dynamic structure factor spectra obtained from molecular dynamics. 
In panel a) the longitudinal phase velocity data are compared to those calculated 
from the corresponding transverse ones (red circles) assuming a frequency-independent 
bulk modulus $B$=59. Within error-bars, we can conclude that the frequency-dependence 
of the longitudinal phase velocity directly reflects that of the transverse one. 
In panel b) the longitudinal excitations broadening data are compared to the corresponding 
transverse ones (red circles, right axis): the two sets of data can be convincingly 
scaled one on top of the other. Within error-bars, we can again conclude that the 
frequency-dependence of the longitudinal data directly reflects that of the 
transverse ones. 
}
\label{Fig3}
\end{figure}

A similar scenario holds for the longitudinal acoustic-like modes. The 
longitudinal sound velocity data (blue squares) are reported in Fig.~3a, and show again 
a decrease at low frequency followed by a positive dispersion for $\omega > 0.8$.
The similarity with the behaviour of the transverse data suggests a common origin
for both. Indeed, the longitudinal and transverse sound velocities in an 
isotropic elastic medium are simply related by the expression:
\begin{equation}
c_L(\omega)=\sqrt{\frac{B(\omega)}{\rho} + \frac{4}{3} c_T^2(\omega)},
\label{Eq2}
\end{equation}
where $\rho$ is the mass density and $B$ is the bulk modulus. 
This equation simply tells that in a glass the shear modulus, $G = \rho c_T^2$, 
the longitudinal modulus, $M = \rho c_L^2$, and the bulk modulus $B$ are 
related, and only two of them give independent information. The low-frequency, 
macroscopic value for the bulk modulus $B(\omega \rightarrow 0)=$59 can be obtained 
from the low-frequency data for $c_L(\omega)$ and $c_T(\omega)$, and is in good 
agreement with a literature value obtained for a slightly different 
system~\cite{leonforte05}. Assuming a frequency independent value for $B$,
we can then estimate the longitudinal sound velocity using the transverse data 
reported in Fig.~2a. The results of this calculation are shown in Fig.~3a 
(red circles).
The good correspondence between the two sets of longitudinal velocity data 
derived directly from $S_L(q,\omega)$ or from $S_T(q,\omega)$ via Eq.~\ref{Eq2} 
strongly supports a scenario where: i) the bulk modulus is frequency independent and 
ii) the frequency dependence of the longitudinal sound velocity simply derives from 
that of the transverse one. This picture is further reinforced by the comparison shown 
in Fig.~3b between the broadening of the longitudinal acoustic-like excitations 
(blue squares, left axis) and that of the transverse ones (red circles, right axis). It 
is clear here that these two quantities, within error-bars, can be scaled one on top of 
the other. This confirms that, within the accuracy of our calculation, the frequency 
dependence that characterizes the longitudinal acoustic-like excitations comes into play 
through the shear component of the longitudinal response, while the bulk component is 
a mere spectator. 

The observation of the $\omega^4$ regime at low frequency in the acoustic attenuation for 
both polarizations is an interesting result and is predicted by several 
models~\cite{klemens51,akkermans85,gurevich03,schirmacher06}.
The simplest one can be formulated in terms of Rayleigh scattering 
of the acoustic excitations from some kind of structural disorder~\cite{klemens51}.
In order to make this argument quantitative one needs to identify the
scattering units that produce it, a difficult task that has lead to different 
conclusions~\cite{elliott92}. Other approaches include a description of the 
disorder in terms of a continuum model with randomly fluctuating 
transverse elastic constants~\cite{schirmacher06,schirmacher07}, or of lattice models 
with disorder in the elastic constants~\cite{schirmacher98}. 
On this basis, it has been quite puzzling not to observe this strong scattering 
regime in previous molecular dynamics 
studies~\cite{mazzacurati96,taraskin99,angelani00,ruocco00,horbach01,schober04}.
It has for example been argued that, in contrast with lattice based models, this 
feature would be absent in realistic glasses due to their intrinsic high level of 
frustration (large internal stress)~\cite{angelani00}. 
The present results solve the issue, clarifying that a strong scattering 
regime is present in fact in realistic structural glasses also, in agreement with 
a number of theoretical~\cite{klemens51,akkermans85,schirmacher98,gurevich03,schirmacher06} 
and experimental~\cite{dietsche79,ruffle03,ruffle06,monaco09} results.
We believe that the reason why this regime got unnoticed before has to do with the fact 
that, beside the obvious cases of too small simulation boxes, the attention was usually 
focused on the longitudinal excitations (the only ones experimentally accessible) where 
this regime is difficult to study in detail even with a box as large as the one used here, 
see Fig.~3b. The present results in fact clarify that the signature of strong scattering 
appears for both polarizations at the same frequency which implies that -- as the 
longitudinal speed of sound is larger than the transverse one (by a factor 2.3 
in the present case) -- the strong scattering regime appears for the longitudinal polarization 
at lower $q$'s than for the transverse one: its observation is then definitively less favorable 
for the longitudinal polarization.
\begin{figure}[t]
\centering
\includegraphics[width=0.48\textwidth]{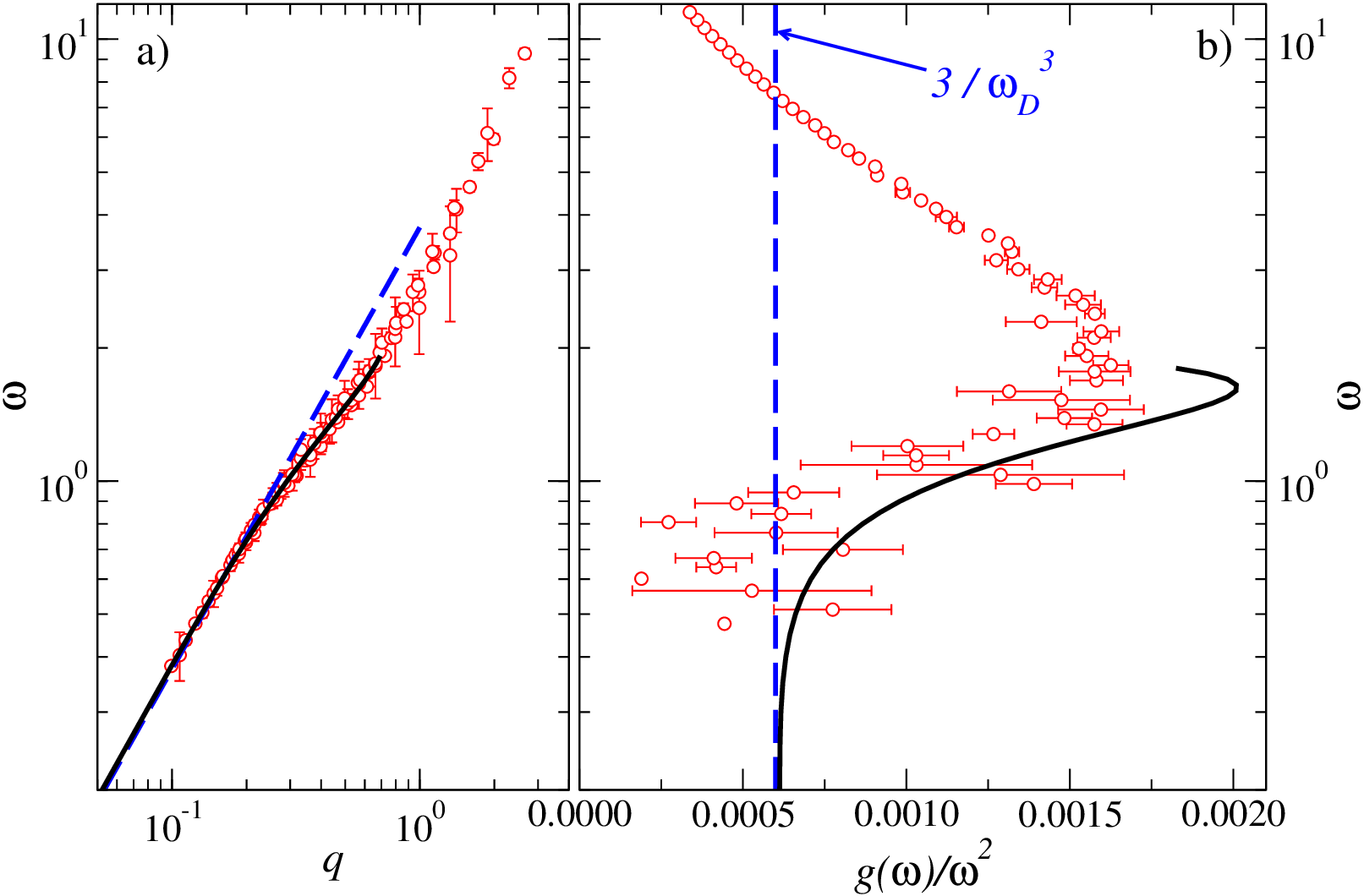}
\caption{
Results of a normal modes analysis of a Lennard-Jones glass in its inherent structures. 
a) $q$-dependence of the lowest frequency transverse eigenvalues of the studied LJ glass 
(circles) together with the function used to empirically describe it (full line). 
The initial slope of the curve corresponds to the $q \rightarrow$0 limit of the 
transverse phase velocity of Fig.~2a (dashed line).
Note that here $q$ is only an approximate quantum number, and that it is increasingly 
difficult to associate higher frequency eigenvalues to specific $q$ values. 
b) Reduced density of states of the studied LJ glass (circles) showing the boson 
peak at $\omega\simeq$~2. The macroscopic, Debye limit (dashed line) is indicated 
as well. The reduced density of states directly obtained from the data in Fig.~4a via 
Eq.~\ref{Eq3} is reported (full line) up to the maximum frequency where this analysis 
is appropriate. The agreement between the two calculations is good -- note that there is no 
adjustable parameter in this comparison. 
This implies that the density of states, up to the boson peak, is well described
by the low-$q$ inflection observed in Fig.~4a. 
}
\label{Fig4}
\end{figure}

More in general, the present results clearly show that the macroscopic and microscopic 
regimes for the acoustic-like excitations are connected by a crossover region where the 
acoustic-like dispersion curves show a considerable 
deformation with respect to a simple linear dependence: the sound velocities decrease 
with increasing frequency thus directly testifying the existence of an abrupt 
breakdown of the Debye approximation in glasses~\cite{monaco09}. This comes together with 
the signature of strong scattering for the acoustic-like excitations across which 
reasonably well defined plane waves transform into a complex pattern of atomic 
vibrations. The connection to the boson peak is also clear: a softening of the 
transverse and longitudinal sound velocities implies an excess in the reduced 
vibrational density of states above the Debye level. Since this softening directly 
appears in the range where the boson peak is observed, or at the corresponding 
temperatures where the excess in the specific heat universally appears in 
glasses~\cite{phillips81}, we can conclude that these anomalies must have 
an acoustic contribution. In the following we make this connection quantitative.

In order to gain further insight into these molecular dynamics results, we performed 
a standard normal modes analysis of the glass in its inherent structures, i.e. we 
have diagonalized the dynamical matrix calculated in local minima of the potential 
energy landscape and derived eigenvalues and eigenstates. The lowest frequency 
eigenstates of transverse polarization are close to be plane waves: it is possible 
to associate them with a leading $q$ value~\cite{leonforte05} (see  Appendix~\ref{appendix-methods}
for details). In Fig.~4a we report the corresponding eigenvalues as a function of $q$ (circles); 
the reduced vibrational density of states derived from the complete normal modes analysis 
is shown in Fig.~4b (circles). 

Fig.~4a corresponds to a pseudo-dispersion curve: it is not formally a dispersion curve 
but only an approximate one since in a disordered system $q$ is not a good quantum number, 
though one can still imagine that it is a reasonable mean to count the modes of low enough 
frequency.
This pseudo-dispersion curve must have a slope for $q \rightarrow$0 that corresponds to the 
transverse velocity. In fact, as shown in Fig.~4a, this slope well corresponds to the 
$q \rightarrow$0 limit of the transverse velocity (dashed line) derived from Fig.~2. 
On increasing $q$, the data in Fig.~4a show an early departure from linearity 
and a clear inflection at $\omega\simeq$~2 which is the frequency where the boson
peak appears (Fig.~4b). This inflection has been interpreted in terms of the failure 
of the classical Born approximation for the description of the continuum elasticity in 
disordered systems~\cite{leonforte05}. It appears at the same frequency where the molecular 
dynamics results for the phase velocities in Fig.~2 and Fig.~3 show a softening. 
This is not surprising as the acoustic-like frequencies can indeed be regarded as average 
values over a distribution of eigenvalues related to the broadening of the Brillouin peaks.
In other words, the velocity decrease observed at frequencies around the boson peak position in 
the acoustic-like excitations of transverse polarization (Fig.~2a) and, through the shear component 
of the longitudinal modulus, in those of longitudinal polarization (Fig.~3a), directly reflects 
an inflection that appears in the $q$-dependence of the low-$q$ transverse eigenvalues.

As far as we can consider $q$ a reasonable quantum number, it is easy to directly estimate 
the vibrational density of states starting from the knowledge of the dispersion curves. 
In a simple plane-wave approach:
\begin{eqnarray}
\frac{g(\omega)}{\omega^2}&=&\frac{1}{q_D^3} \Bigl[\Bigl(\frac{q^2}{\omega^2}
\frac{\partial q}{\partial \omega}\Bigr)_L + 2 \Bigl(\frac{q^2}{\omega^2} 
\frac{\partial q}{\partial \omega} \Bigr)_T \Bigr] \nonumber\\
& \simeq& \frac{3}{q_D^3}
\Bigl(\frac{c_T}{c_D}\Bigr)^3 
\Bigl(\frac{q^2}{\omega^2} \frac{\partial q}{\partial \omega} \Bigr)_T , 
\label{Eq3}
\end{eqnarray}
where $L$ and $T$ stand for the longitudinal and transverse polarization, and $c_D$ is the Debye 
velocity. The approximation 
in Eq.~3 is justified by the fact that the longitudinal contribution 
to the total vibrational density of states is small since it scales with 
$c_T^3 / (2 c_L^3)$=4\%. It is clear that in order to follow this approach we shall 
use the pseudo-dispersion curve obtained from the normal modes analysis since in that case the 
low-frequency eigenvalues are reasonably close to being plane waves. In order 
to compute Eq.~\ref{Eq3}, instead of directly differentiating the dispersion curve, 
we choose to use an empirical model to fit the simulation data 
(full line through the points in Fig.~4a) and then to differentiate the model function. 
The result of this calculation is shown in Fig.~4b (full line) and, as it can be 
appreciated, it well describes the boson peak. It is worth underlining that 
no additional input but the knowledge of the dispersion curve is required to perform this 
calculation. It is then clear that the reduced density of states of the studied glass, up to 
the boson peak, originates from a deformation of the pseudo-dispersion curve of eigenstates 
that can still be approximately described using plane waves.
\begin{figure}[t]
\centering
\includegraphics[width=0.48\textwidth]{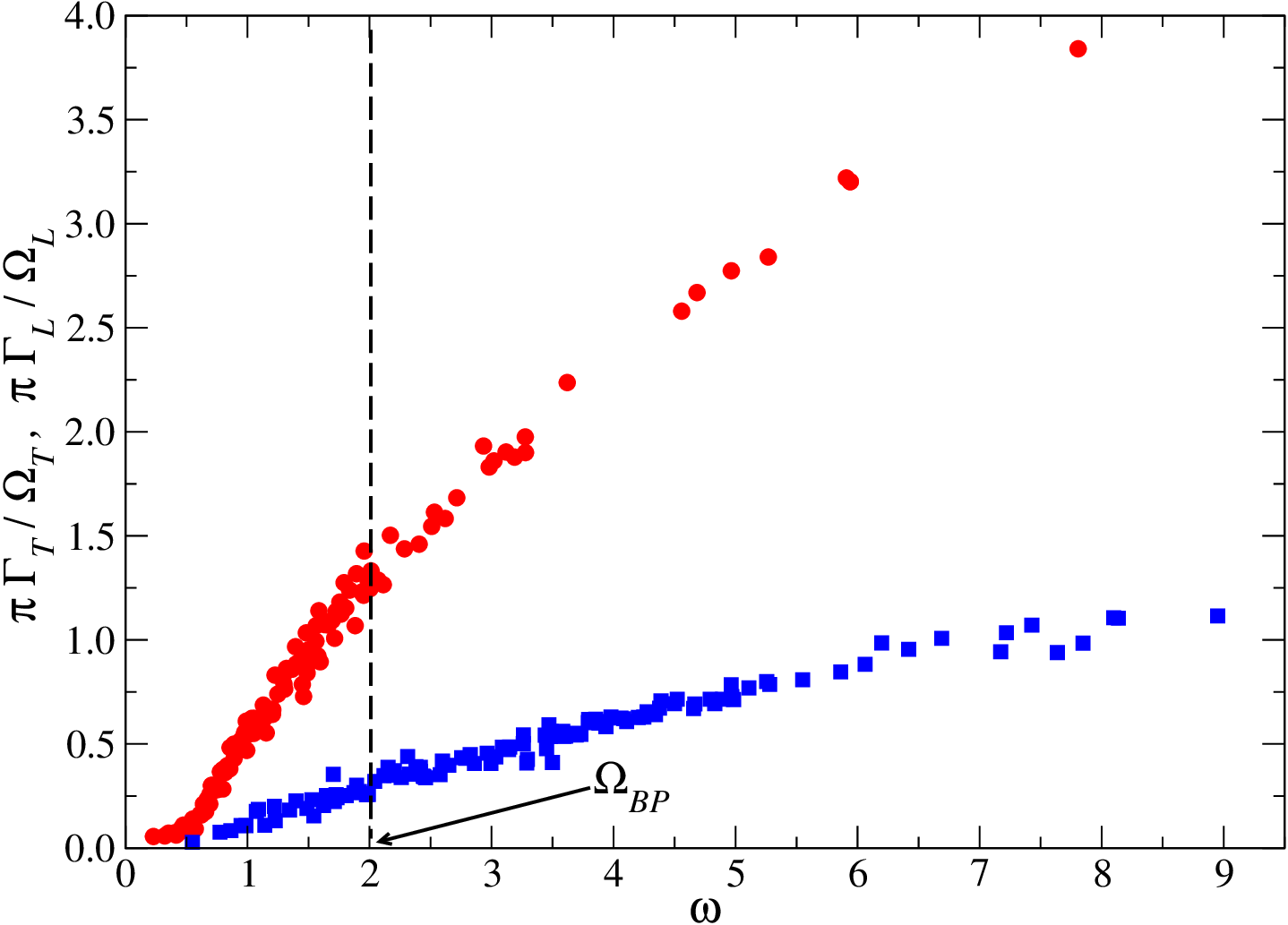}
\caption{
The ratio $\pi \Gamma / \Omega$ is reported as a function of frequency for both 
the transverse (circles) and longitudinal (squares) polarization. 
The frequency where this ratio equals one defines the Ioffe-Regel limit. For the 
transverse case, this limit falls very close to the boson peak position, indicated 
by the vertical dashed line. Different is the case of the longitudinal polarization, 
where the Ioffe-Regel limit appears at a frequency $\sim$~3 times higher than the boson 
peak position.
}
\label{Fig5}
\end{figure}

The molecular dynamics results presented above also allow us to calculate the 
Ioffe-Regel limit for the acoustic-like excitations, defined as the frequency where 
$\Omega=\pi \Gamma$ for both polarizations~\cite{ruffle06}.
This limit corresponds to the frequency where 
the decay time of the acoustic-like excitations first matches half of the
corresponding oscillation period, and therefore marks 
somehow an extreme upper bound in frequency for the validity of a plane-waves approach as 
a starting point to describe the acoustic-like excitations. 
As shown in Fig.~5, the present data confirm that the Ioffe-Regel limit for 
the transverse excitations is located close to the boson peak 
position~\cite{schober04,shintani08}. 
Fig.~5 also shows that the Ioffe-Regel limit is reached at different 
frequencies for the longitudinal and transverse excitations, and that the longitudinal 
one shows up at a frequency higher than the boson peak position~\cite{shintani08}. 
However, it is important to underline that this last result seems not to be general: 
in a simulation study of a silica glass the Ioffe-Regel crossover was found to appear 
at the same frequency for both polarizations~\cite{taraskin99}.
The connection between the Ioffe-Regel limit for the transverse excitations and the 
boson peak~\cite{ruffle06,shintani08} is clarified by the discussion above, which 
shows that this limit is reached in fact in the frequency range where the continuum 
approximation for the acoustic excitations breaks down.
\section{Conclusions}
Summing up, the present simulation results shed new light on the well known universal 
anomaly observed in the specific heat of glasses in the $T \sim$~10~K temperature range 
and related to the boson peak in the vibrational density of states at frequencies of 
$\sim$1~THz. We have shown that in glasses the elastic continuum approximation for the 
acoustic-like excitations breaks down abruptly on the mesoscopic, medium-range-order 
length-scale of about ten interatomic spacings, where it still works well for the 
corresponding crystalline systems. 
This breakdown is signaled by a deformation of the pseudo-dispersion curve and 
corresponds to a marked reduction of the sound velocity on the mesoscopic scale. 
This turns out to be the closely related to the aforementioned anomalies in the 
specific heat and vibrational density of states, that can be finally traced back onto 
elastic properties specific to glasses. 

Finally, in order to put the present results in some perspective and in connection 
with experiments, it is important to emphasize once more that they refer to a simple 
monatomic LJ glass quenched with a cooling rate out of reach experimentally -- in 
other words there is no experimental analogue of the glass studied here. 
Still, we believe that such a simple model system has the great advantage of 
clearly grasping fundamental features that -- though observed in bits and pieces in many
experiments -- are often hidden by a number of additional effects in real glasses. 
This -- we believe -- is the reason why the vibrational properties
of glasses are still a debated issue after several decades of studies and discussions. 
For example, the high-frequency vibrational dynamics of the LJ monatomic glass has been 
proven to be well described within the harmonic approximation~\cite{mazzacurati96,ruocco00}. 
However, clearly real glasses are anharmonic systems -- in what cases will anharmonicity 
start to play a role there? Moreover, differently from the investigated model, real glasses 
are often characterized by the presence of intramolecular or optic-like modes -- this issue 
is of course (and on purpose) completely disregarded here.
All of these questions will require studies on further and more complex models in order to 
be fully addressed.
\begin{acknowledgments}
We thank A.~Tanguy for discussions.
\end{acknowledgments}
\appendix
\section{Numerical methods}
\label{appendix-methods}
The present numerical investigation has been performed using simulation boxes 
of different sizes containing up to $N\sim$10$^7$ atoms, interacting via a 
Lennard-Jones potential with a cutoff $r_c =$2.5 and periodic boundary conditions.
This has been realized using the large scale massively parallel molecular 
dynamics computer simulation code LAMMPS~\cite{plimpton95}.
A standard microcanonical classical molecular dynamics simulation, 
carried out at the constant number density $\widehat{\rho}=$~1.015 and at temperature 
$T=$~2 in the normal liquid phase is followed by a fast quench 
($dT/dt\sim$4$\times$10$^2$) down to $T$=~10$^{-3}$. 
The quenched glass sample is relaxed for a time (dependent on the sample size) 
sufficient to have a constant total energy. The atomic positions, $\mathbf{r}_i(t)$, 
and velocities, $\mathbf{v}_i(t)$, have then been stored for a time 
(again dependent on the sample size) sufficient to get the desired resolution function. 
The time correlation functions required to obtain the dynamic structure factor, 
$S_L(q,\omega)$, and its analogous function for the transverse excitations, 
$S_T(q,\omega)$ have been computed as:
\begin{equation}
S_\alpha(q,\omega)=\frac{1}{2 \pi N}\Bigl(\frac{q}{\omega}\Bigr)^2\int dt\;
\langle \mathbf{j}_{\alpha}(q,t)\cdot\mathbf{j}^{\dagger}_{\alpha}(q,0)\rangle
e^{i\omega t},
\label{Eq4}
\end{equation}
where $\alpha$ is $L$ or $T$ and:
\begin{eqnarray}
\mathbf{j}_{L}(q,t)&=&\sum_{i=1}^{N}\left[\mathbf{v}_i(t)\cdot
\widehat{\mathbf{q}}\right]\widehat{\mathbf{q}} \;
e^{i \mathbf{q}\cdot\mathbf{r}_i(t)}, \\
\mathbf{j}_{T}(q,t)&=& \sum_{i=1}^{N}\left\{
\left[\mathbf{v}_i(t)\cdot \widehat{\mathbf{q}}\right]
\widehat{\mathbf{q}} - \mathbf{v}_i(t) \right\}
\;e^{i \mathbf{q}\cdot\mathbf{r}_i(t)},
\label{Eq5}
\end{eqnarray}
with $\widehat{\mathbf{q}}=\mathbf{q}/\vert\mathbf{q}\vert$. 

Addressing the issue of the dependence of the obtained results on the quenching
rate is a difficult task for the studied system due to the fact that it easily 
crystallizes. The smallest quenching rate compatible with the long simulation times 
needed to reach the desired resolution and with the large sizes of the simulation 
boxes is about 5 decades smaller ($dT/dt\sim$~4$\times$10$^{-3}$) than the one used 
for this study. We have checked that the molecular dynamics results reported here
are independent of the quenching rate in this range.

For the glass studied here a standard normal modes analysis has been carried out to 
derive the vibrational density of states, $g(\omega)$, from the eigenvalues of the 
dynamical matrix calculated in the inherent structures of the glass. 
Moreover, the pseudo-dispersion curve for the transverse eigenvalues of the system 
reported in Fig.~4a has been constructed following Ref.~\cite{leonforte05}. More in 
detail, for each of the considered simulation boxes, we collected the four lowest frequency 
degenerate eigenvalues obtained from the diagonalization of the dynamical matrix. 
These modes correspond to the largest wavelength standing waves for the simulation box. 
Since the transverse sound velocity is $\simeq$2.3 times 
smaller than the longitudinal one, these four degenerate eigenvalues are all of transverse 
polarization. For example, the first eigenvalue has a degeneracy of 12 and can be associated 
to wavevectors of the ($\pm$1,0,0) family; the second eigenvalue has a degeneracy of 24 
and can be associated to the wavevectors of the ($\pm$1,$\pm$1,0) family, and so on. 
These eigenvalues are size dependent: performing this analysis on simulation boxes of larger 
and larger size (up to N=256000 particles), we then selected transverse eigenvalues with 
smaller and smaller $q$'s. A quite broad range of frequencies and $q$'s could thus be explored.
This procedure can be expected to lead to reasonable results only as far as the degeneracy of 
the eigenvalues shows up clearly enough to suggest a one-to-one relation to the corresponding 
values for $q$, and becomes less and less reliable on decreasing the box size or on increasing 
$q$. In the present case, we find that the highest $q$ values up to which this analysis is 
still reasonable are $q \sim$~0.2-0.3, corresponding to $\omega \sim$~2, i.e. basically up to 
the boson peak position, and becomes less and less reliable on further increasing $q$.


\begin{thebibliography}{99}
%
\bibitem{shapiro66}
Shapiro S~M, Gammon R~W, Cummins H~Z (1966) Brillouin scattering 
spectra of crystalline quartz, fused quartz and glass. 
{\it Appl. Phys. Lett.} 9:157-159.

\bibitem{sette98}
Sette F, Krisch M, Masciovecchio C, Ruocco G, Monaco G (1998) Dynamics of 
glasses and glass-forming liquids studied by inelastic x-ray scattering.
{\it Science} 280:1550-1555.

\bibitem{bove05}
Bove L~E, et al. (2005) Brillouin neutron scattering of v-GeO$_2$.
{\it Europhys. Lett.} 71:563-569.

\bibitem{rahman76}
Rahman A, Mandell M~J, McTague J~P (1976) Molecular dynamics study of an 
amorphous Lennard-Jones system at low temperature.
{\it J. Chem. Phys.} 64:1564-1568.

\bibitem{grest82}
Grest G~S, Nagel S~R, Rahman A (1982) Longitudinal and transverse excitations in a glass.
{\it Phys. Rev. Lett.} 49:1271-1274.

\bibitem{mazzacurati96}
Mazzacurati V, Ruocco G, Sampoli M (1996) Low-frequency atomic motion in a model glass.
{\it Europhys. Lett.} 34:681-686.

\bibitem{taraskin99}
Taraskin S~N, Elliott S~R (1999) Low frequency vibrational excitations in vitreous silica: 
the Ioffe-Regel limit. {\it J. Phys. Condens. Matter} 11:A219-A227.

\bibitem{angelani00}
Angelani L, Montagna M, Ruocco G, Viliani G (2000) Frustration and sound
attenuation in structural glasses. {\it Phys. Rev. Lett.} 84:4874-4877.

\bibitem{ruocco00}
Ruocco G, et al. (2000) Relaxation processes in harmonic glasses?
{\it Phys. Rev. Lett.} 84:5788-5791.

\bibitem{horbach01}
Horbach J, Kob W, Binder K (2001) High frequency sound and the boson peak in amorphous silica.
{\it Eur. Phys. J. B} 19:531-543.

\bibitem{schober04}
Schober H.R. (2004) Vibrations and relaxations in a soft sphere glass: boson peak and structure factors.
{\it J. Phys. Condens. Matter} 16:S2659-S2670.

\bibitem{grest84}
Grest G~S, Nagel S~R, Rahman A (1982) Zone boundaries in glasses.
{\it Phys. Rev. B} R29:5968-5971.

\bibitem{phillips81} 
Phillips W~A, ed (1981) Amorphous solids: low temperature properties (Springer, Berlin).

\bibitem{karpov83}
Karpov V~G, Klinger M~I, Ignat'ev F~N (1983) Theory of the 
low-temperature anomalies in the thermal properties of amorphous 
structures. {\it Sov. Phys. JETP} 57:439-448.

\bibitem{akkermans85}
Akkermans E, Maynard R (1985) Weak localization and anharmonicity of phonons. 
{\it Phys. Rev. B} 32:7850-7862.

\bibitem{dove97}
Dove M, et al. (1997) 
Floppy modes in crystalline and amorphous silicates.
{\it Phys. Rev. Lett.} 78:1070-1073.

\bibitem{schirmacher98}
Schirmacher W, Diezemann G, Ganter C (1998) Harmonic 
vibrational excitations in disordered solids and the "boson 
peak". {\it Phys. Rev. Lett.} 81:136-139.

\bibitem{hehlen00}
Hehlen B, et al. (2000) Hyper-Raman scattering observation of 
the boson peak in vitreous silica. {\it Phys. Rev. Lett.} 84:5355-5358.

\bibitem{taraskin01}
Taraskin S~N, Loh Y~L, Natarajan G, Elliott S~R (2001) 
Origin of the boson peak in systems with lattice disorder.
{\it Phys. Rev. Lett.} 86:1255-1258.

\bibitem{wittmer02}
Wittmer J~P, Tanguy A, Barrat J-L, Lewis L (2002) 
Vibrations of amorphous, nanometric structures: when does continuum theory apply?
{\it Europhys. Lett.} 57:423-429.

\bibitem{gurevich03}
Gurevich V~L, Parshin D~A, Schober H~R (2003) Anharmonicity, 
vibrational instability, and the Boson peak in glasses. 
{\it Phys. Rev. B} 67:094203.

\bibitem{lubchenko03}
Lubchenko V, Wolynes P~G (2003) 
The origin of the boson peak and thermal conductivity plateau in low-temperature glasses. 
{\it Proc. Natl. Acad. Sci. USA} 100:1515-1518.

\bibitem{grigera03}
Grigera T~S, Mart\'in-Mayor V, Parisi G, Verrocchio P (2003) 
Phonon interpretation of the "boson peak" in supercooled liquids.
{\it Nature} 422:289-292.

\bibitem{leonforte05}
Leonforte F, Boissi\'ere R, Tanguy A, Wittmer J~P, Barrat J-L (2005) 
Continuum limit of amorphous elastic bodies. III. Three-dimensional systems.
{\it Phys. Rev. B} 72:224206.

\bibitem{schirmacher06}
Schirmacher W (2006) Thermal conductivity of glassy materials 
and the "boson peak". {\it Europhys. Lett.} 73:892-898.

\bibitem{schirmacher07}
Schirmacher W, Ruocco G, Scopigno T (2007) Acoustic attenuation in glasses 
and its relation with the boson peak. {\it Phys. Rev. Lett.} 98:025501.

\bibitem{shintani08}
Shintani H, Tanaka H (2008)
Universal link between the boson peak and transverse phonons in glass. 
{\it Nature Mater.} 7:870-877.

\bibitem{buchenau84}
Buchenau U, N\"ucker N, Dianoux A~J (1984) 
Neutron scattering study of the low frequency vibrations in vitreous silica.
{\it Phys. Rev. Lett.} 53:2316-2319.

\bibitem{malinovsky91}
Malinovsky V~K, Novikov V~N, Sokolov A~P (1991) Log-normal 
spectrum of low-energy vibrational excitations in glasses.
{\it Phys. Lett. A} 153:63-66.

\bibitem{rothenfusser83}
Rothenfusser M, Dietsche W, Kinder H (1983) 
Linear dispersion of transverse high-frequency phonons in vitreous silica.
{\it Phys. Rev. B} 27:5196-5198.

\bibitem{monaco09}
Monaco G, Giordano V (2009) 
Breakdown of the Debye approximation for the acoustic modes with nanometric wavelengths in glasses. 
{\it Proc. Natl. Acad. Sci. USA} 106:3659-3663. 

\bibitem{ruffle03}
Ruffl\'e B, Foret M, Courtens E, Vacher R, Monaco G (2003) 
Observation of the onset of strong scattering on high frequency 
acoustic phonons in densified silica glass.
{\it Phys. Rev. Lett.} 90:095502.

\bibitem{ruffle06}
Ruffl\'e B, Guimbr\`etiere G, Courtens E, Vacher R, Monaco G (2006) 
Glass-specific behavior in the damping of acousticlike vibrations.
{\it Phys. Rev. Lett.} 96:045502.

\bibitem{dietsche79}
Dietsche W, Kinder H (1979) 
Spectroscopy of phonon scattering in glass.
{\it Phys. Rev. Lett.} 43:1413-1417.

\bibitem{masciovecchio06}
Masciovecchio C, et al. (2006) 
Evidence for a crossover in the frequency dependence
of the acoustic attenuation in vitreous silica.
{\it Phys. Rev. Lett.} 97:035501.

\bibitem{robles03}
Robles M, L\'opez de Haro M (2003)
On the liquid-glass transition line in monatomic Lennard-Jones fluids.
{\it Europhys. Lett.} 62:56-62.

\bibitem{klemens51}
Klemens P~G (1951) 
The thermal conductivity of dielectric solids at low temperatures. 
{\it Proc. Roy. Soc. A (London)} 208:108-133.

\bibitem{elliott92}
Elliott S~R (1992) 
A unified model for the low-energy vibrational behaviour 
of amorphous solids.
{\it Europhys. Lett.} 19:201-206.

\bibitem{plimpton95}
Plimpton S~J (1995)
Fast parallel algorithms for short-range molecular dynamics.
{\it J. Comp. Phys.} 117:1-19.
%
\end{thebibliography}
\end{document}